\documentclass[twocolumn]{aastex631}
\usepackage{lineno}

\begin{document}

\title{Detection of X-ray polarized emission and accretion-disk winds with {\it IXPE} and {\it NICER} in the black-hole X-ray binary 4U 1630--47}
\shorttitle{X-ray polarimetry of the BHC 4U 1630--47}

\author{Divya Rawat}
\affiliation{Inter-University Center for Astronomy and Astrophysics, Ganeshkhind, Pune 411007, India}
\thanks{E-mail: rawatdivya838@gmail.com (DR)}
\author{Akash Garg}
\affiliation{Inter-University Center for Astronomy and Astrophysics, Ganeshkhind, Pune 411007, India}
\author{Mariano M\'endez}
\affiliation{Kapteyn Astronomical Institute, University of Groningen, PO BOX 800, Groningen NL-9700 AV, the Netherlands}

\begin{abstract}
We detect a high level of polarization in the X-ray emission of the black-hole binary 4U 1630--47 in an observation with the {\it Imaging X-ray Polarimetry Explorer}. The $2-8$ keV polarization degree is 8 \% at a position angle of $18^\circ$, with the polarization degree increasing significantly with energy, from $\sim 6$ \% at $\sim 2$ keV to $\sim 11$ \% at $\sim 8$ keV. The continuum emission in the spectrum of simultaneous observations with the {\it Neutron Star Interior Composition Explorer}, {\it NICER}, is well described with only a thermal disc spectrum, with stringent upper limits to any Comptonized emission from the corona. Together with the lack of significant variability in the Fourier power spectrum, this suggests that the source was in the high-soft state at the time of these observations. The {\it NICER} spectrum reveals the presence of several absorption lines in the $6-9$ keV band that we fit with two ionized absorbers, providing evidence of the presence of a strong disk wind, which supports the idea that the source was in the soft state. Previous measurements of X-ray polarization in other sources in harder states were associated with the corona or the jet in those systems. Given that the corona is significantly absent in this observation of 4U 1630--47, and that the jet in black-hole binaries is quenched in the high-soft state, we speculate that in this observation of 4U 1630--47, the polarization likely arises from the direct and reflected radiation of the accretion disk in this source.
\end{abstract}


\section{Introduction} \label{sec:intro}

4U 1630$-$47 is a transient low-mass X-ray binary source discovered in the 1970s with {\it{Uhuru}} and {\it{Ariel V}} \citep{jo76}, and {\it{Vela 5B}} X-ray monitor \citep{pr86}, independently. \cite{ki14} and \cite{li22} reported that the system harbours a highly spinning black hole located at a distance of $\sim 10$ kpc \citep{se14}. From the presence of X-ray dips in the light curve \citep{ku98}, and from fits to the X-ray spectra with a reflection component \citep{di13} a high inclination angle of the system, $i \sim 60^{\circ}$ is reported. X-ray spectral studies show that some of the outbursts are unusually soft for this source, with a lack of hard emission above 30 keV \citep{ca15}. Blue-shifted iron absorption lines in the 6–9 keV band have been reported from this source \citep{ku07,ro14,mi15}. Using Chandra and Swift observation, \citet{ka18} observed that the source is behind a massive molecular cloud, MC -79.\\

Over the last five decades, the spectral and timing capabilities of various X-ray observatories have been exploited to understand the accretion-flow geometry of 4U 1630--47. Still, a clear picture of the disk-corona system has not yet emerged. In this context, polarimetric studies can potentially constrain the geometry of the corona. Using radiative transfer simulations \citet{po23} showed that for edge-on systems, the polarization degree (PD) could exceed 10 \%. Furthermore, if the polarization is due to electron scattering in the accretion disk, the polarization angle would be aligned with the disk plane \citep{ch47,ch60,so63,fa23}. The potentially high inclination angle of the binary system 4U 1630--47 \citep{ku98} makes it an excellent source to explore the properties of the disk and the corona using X-ray polarization measurements. 

In this Letter, we study simultaneous observations of 4U 1630--47 during its 2022 outburst with the {\it Imaging X-ray Polarimetry Explorer}, {\it IXPE} \citep{ma20,we22}, and the {\it Neutron Star Interior Composition Explorer}, {\it NICER}. We describe the observations and data analysis in Section \ref{sec_obs}, the polarimetric and spectro-polarimetric results in Section \ref{Sec_res} and discuss our findings in Section \ref{sec_concl}.

\section{Observation and Data Analysis}\label{sec_obs}
\textit{IXPE} is a small explorer mission launched by NASA in alliance with the Italian Space Agency (ASI) on 2021 December 9. It is the first-ever satellite that measures spatial, temporal, and energy-resolved polarimetric data in a soft X-ray band of $2-8$ keV. The satellite consists of three telescopes, each composed of a focussing Mirror Module Assembly and a Gas Pixel Detector unit at the focus to image the short photoelectric tracks produced as a result of X-ray absorption \citep{ma20,ba21,di22,we22}. \textit{IXPE} observed the black-hole X-ray binary 4U 1630--47 from 2022-08-23 to 2022-09-02 for $\sim$ 460 ks. We analyzed the \textit{IXPE} data using the simulation and analysis framework {\sc{ixpeobssim}} 29.2.0\footnote{\label{note2}\url{https://ixpeobssim.readthedocs.io/en/latest/overview.html}} \citep{ba22}, designed to simulate observations and extract science products from the level2 event files.\\

Using the {\sc{xpselect}} tool within {\sc{ixpeobssim}}, we filtered the level2 event files to obtain cleaned event files for the source and the background separately. We used a circular source region of 60'' and an annular background region of inner and outer radii of 180'' and 240'', respectively, for all three detector units (DU). Further, we applied different binning algorithms using the tool {\sc{xpbin}} of {\sc{ixpeobssim}} to produce images and spectra. Specifically, we used the option {\sc pcube} of the tool {\sc xpbin} to compute a model-independent estimate of the polarization based on the detected photons, and to generate the count, $I$, and Stokes, $Q$ and $U$, spectra for all three DUs. All the science products were produced following the unweighted method. We utilized the response matrices version v012 of {\sc{ixpeobssim}} in
 CALDB version 20170101$\_$02\footnote{\url{https://heasarc.gsfc.nasa.gov/docs/ixpe/caldb/}}. \\
 \begin{figure}
\centering\includegraphics[width=0.54\textwidth,height=0.3\textheight,angle=0]{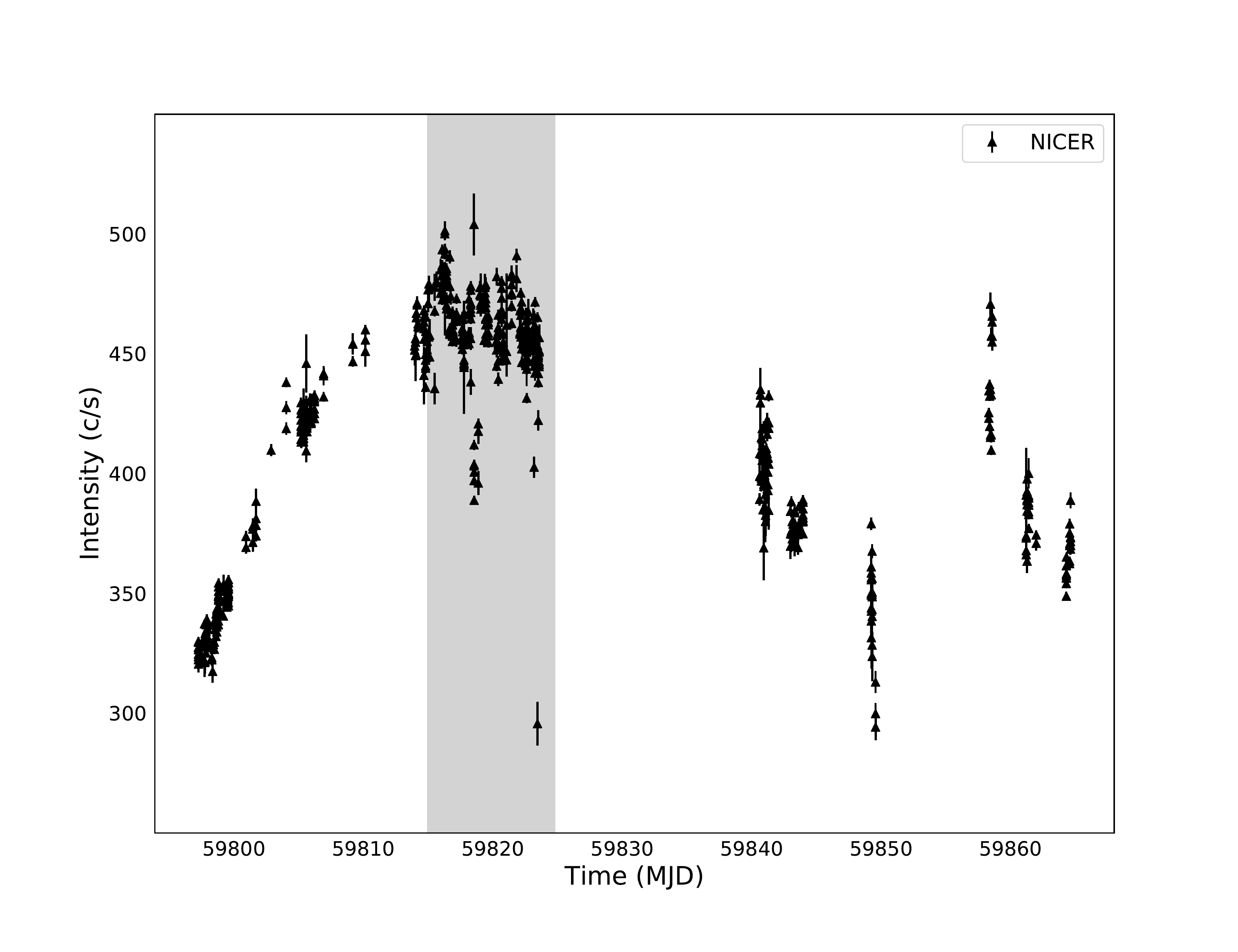}
\centering\includegraphics[width=0.54\textwidth,height=0.3\textheight,angle=0]{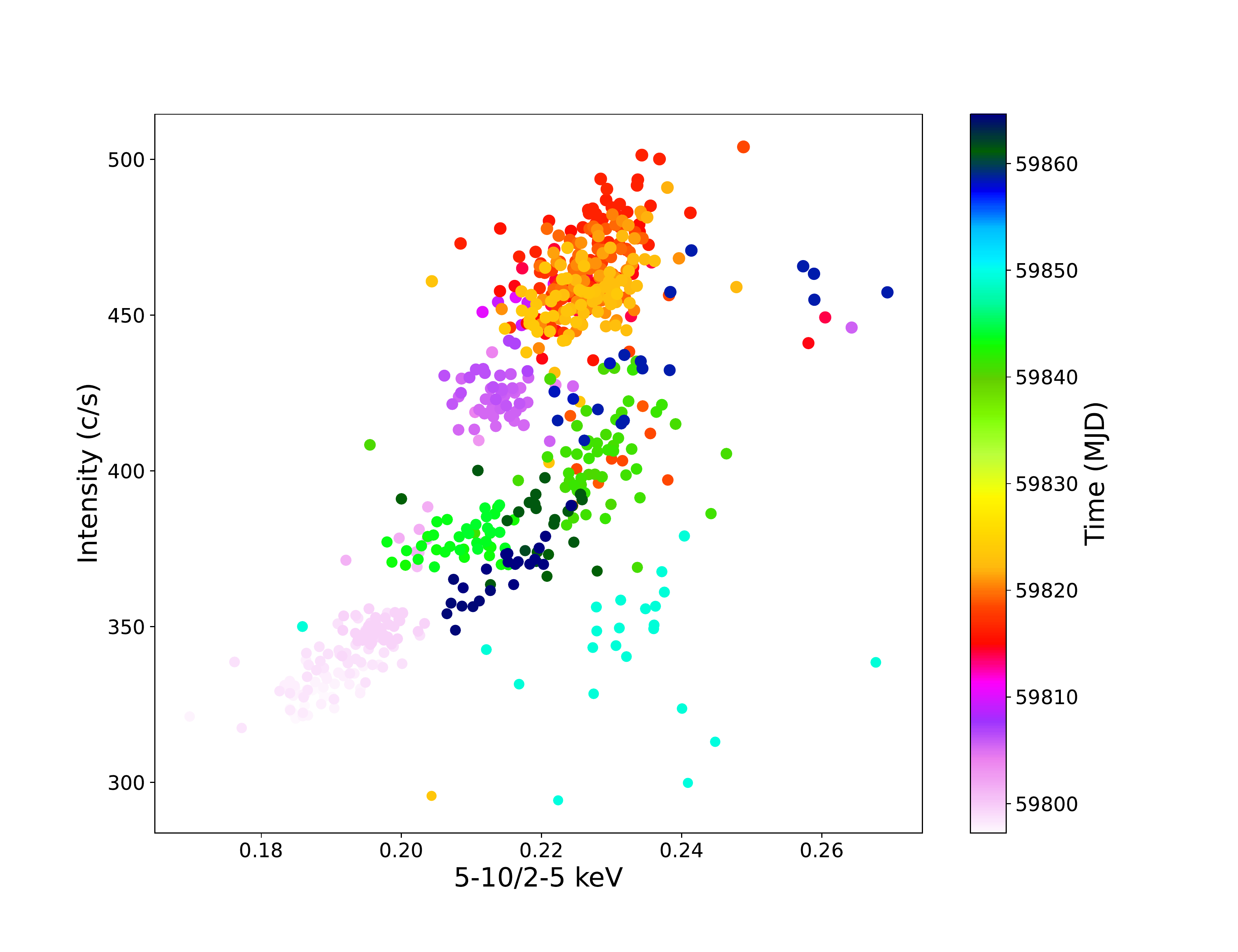}
\caption{Upper panel: {\it{NICER}} light curve of 4U 1630$-$47 in the $0.2-12.0$ keV band. The shaded area represents the  simultaneous {\it{IXPE}}  observation of the source. The bin size of the {\it{NICER}} light curve is 100 s. Lower panel: Hardness-Intensity diagram of 4U 1630$-$47 using NICER data.}
\label{lightcurve}
\end{figure}
\textit{NICER} monitored 4U 1630--47 from 2022-08-22 to 2022-10-12. We have processed the {\it{NICER}} level2 data applying the standard calibration and screening criteria using the {\sc{nicerl2}}\footnote{\url{https://heasarc.gsfc.nasa.gov/docs/nicer/analysis_threads/nicerl2/}} task for each observation. We then merged the clean event files using the {\sc nimpumerge}\footnote{\url{https://heasarc.gsfc.nasa.gov/docs/software/lheasoft/help/nimpumerge.html}} task. We show the 100-s binned \textit{NICER} light curve from 2022-08-22 to 2022-10-12 in the upper panel of Figure \ref{lightcurve}. The shaded area shows the times of the simultaneous {\it{IXPE}} observation of the source. In the lower panel of Figure \ref{lightcurve} we show the hardness-intensity diagram using these same data. In this diagram, the intensity is defined as the background-subtracted count rate in the $0.2-12.0$ keV band, and the hardness is the ratio of the background-subtracted count rates in the $5-10$ keV band to the $2-5$ keV band. The color of the points represents the time in MJD units.\\
For the {\it {NICER}} observations listed in the Appendix Table \ref{obs_log}, we extracted Power Density Spectra (PDS) using the General High-energy Aperiodic Timing Software (GHATS)\footnote{\url{http://www.brera.inaf.it/utenti/belloni/GHATS_Package/Home.html}}. For this, we segmented the 0.2–12.0 keV lightcurve into 1096 intervals of 20.48 s at a time resolution of 2.5$\times$ 10$^{-3}$ s (Nyquist frequency=200 Hz). We averaged all the 1096 PDS to produce a single power spectrum for all observations combined that we rebinned logarithmically. We fitted this PDS with a model consisting of a constant representing the Poisson level and a Lorentzian function representing any noise component from the source (For details of the PDS extraction of NICER data please see the analysis section of \citealt{ra23}). The PDS shows no significant variability, with a 95 \% upper limit of 3.2 \% for the rms amplitude in the $0.05-200$ Hz frequency range.\\

For the spectral analysis, we extracted data from the period in which the observations of the two satellites overlapped. The details of the observations we used, including ObsIDs, start and end times and exposure times, are given in Table \ref{obs_log}. We then extracted the source and background spectra using the \textit{NICER} background estimator tool {\sc{3c50\_rgv7}} \citep{re22}.

We used {\sc{Heasoft}} version 6.31.1\footnote{\url{https://heasarc.gsfc.nasa.gov/lheasoft/download.html}} and {\sc{CALDB}} version 20221001\footnote{\url{https://heasarc.gsfc.nasa.gov/docs/heasarc/caldb/nicer/}} to create the response ({\sc rmf}) and ancillary response ({\sc arf}) files. We added systematic errors to the source spectrum to account for calibration uncertainties using the tool {\sc{niphasyserr}} and rebinned the data to the optimal resolution using the tool {\sc{ftgrouppha}}. To fit the spectra, we used the X-ray spectra fitting package {\sc{xspec version 12.13.0}} \citep{ar96}

\begin{figure}
\centering\includegraphics[scale=0.39,angle=0]{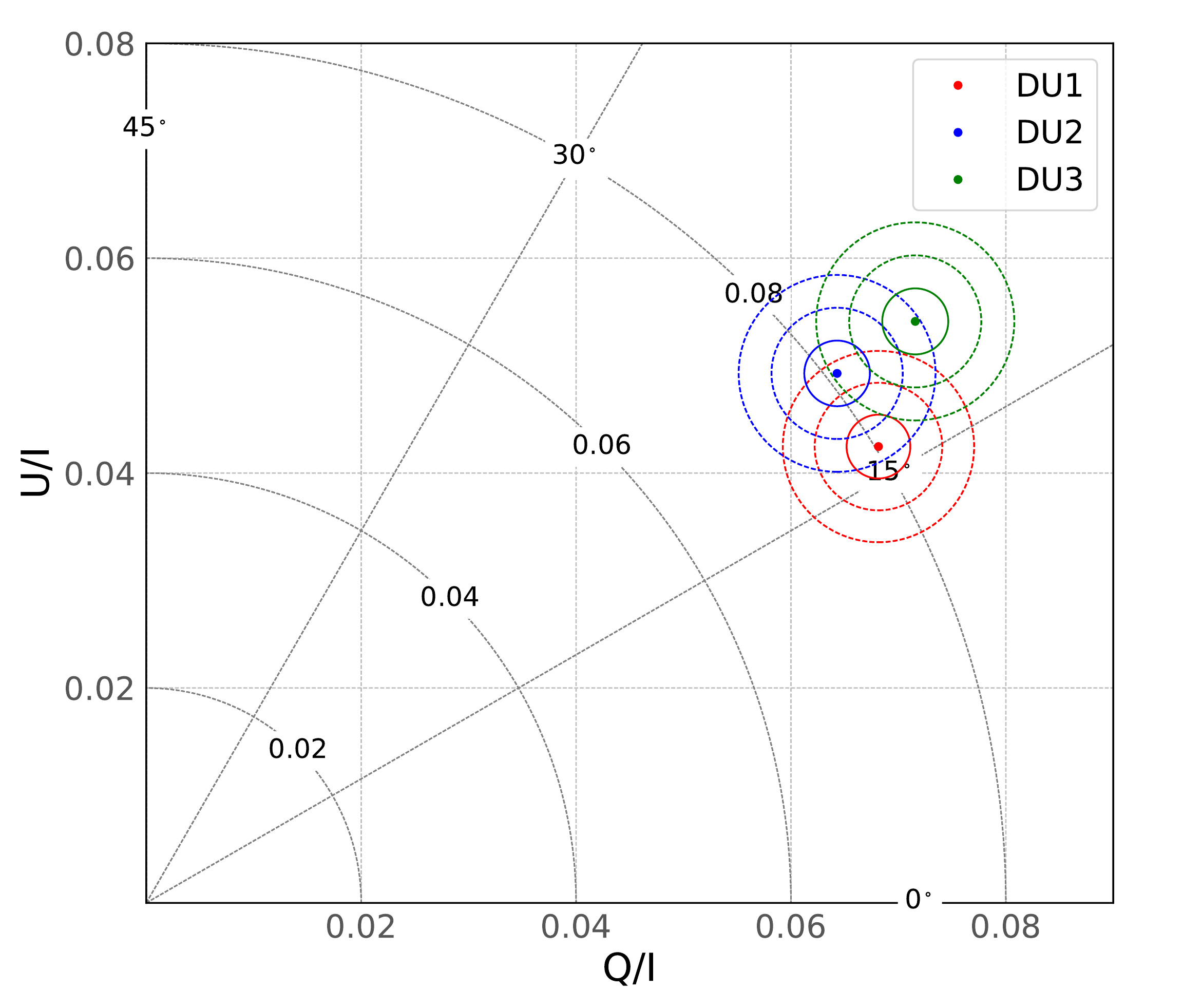}
\caption{The polarization degree, PD, polarization angle, PA, and normalized stokes parameter, $Q/I$ and $U/I$, of 4U 1630$-$47 measured with the three detectors, DU1, DU2, DU3, onboard {\it IXPE}. The measurements were obtained using the {\sc{pcube}} tool in the 2--8 keV band. The contours are at 68.27, 95.45, and 99.73 per cent confidence levels for both Stokes parameters.}
\label{2_8_stoke}
\end{figure}

\begin{figure*}
\centering\includegraphics[scale=0.65,angle=0]{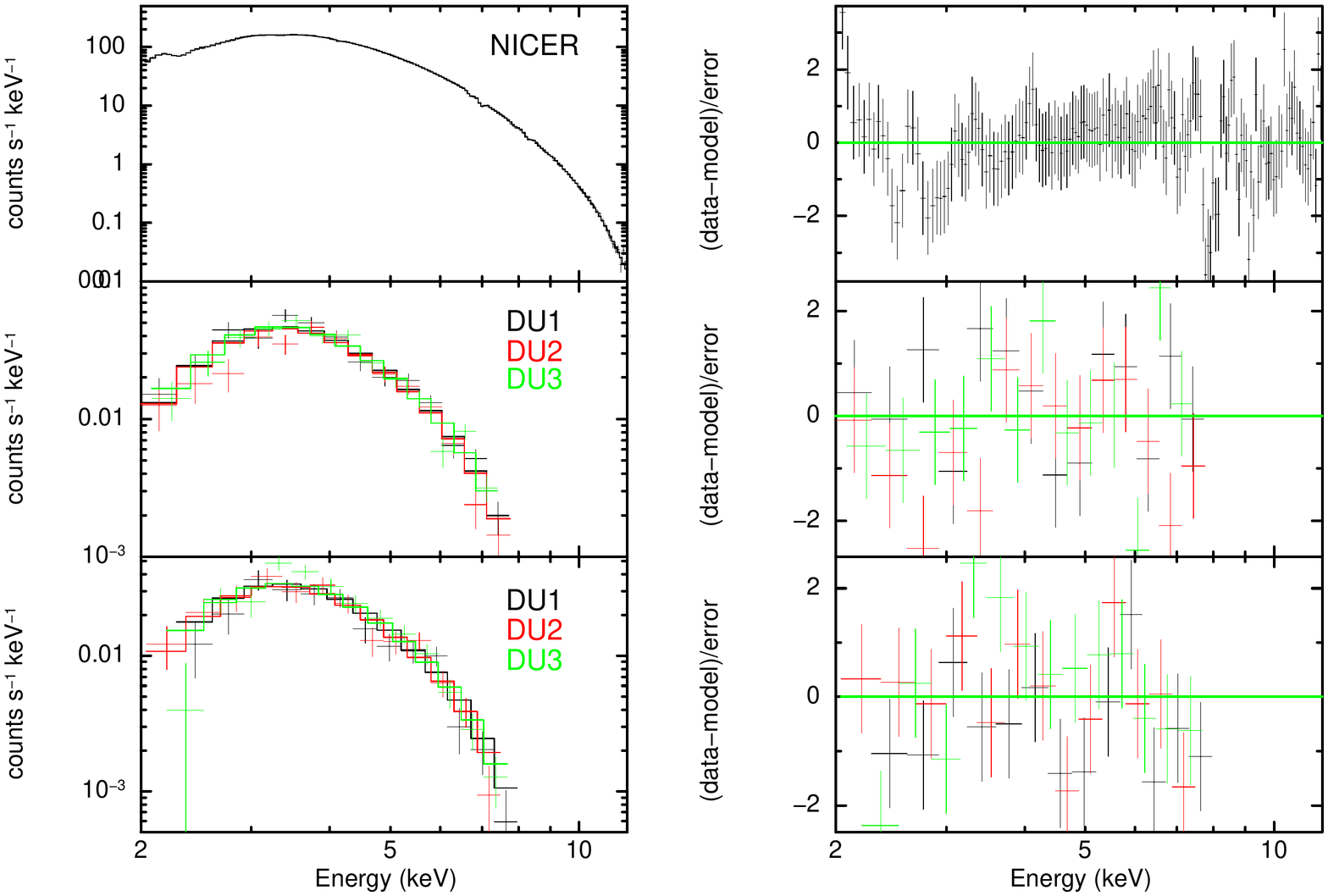}
\caption{Results of the simultaneous fits to the {\it{NICER}} spectra (top left panel) and {\it IXPE} Stokes spectra, $Q$ (middle left panel), and $U$ (bottom left panel) of 4U 1630$-$47 fitted with model  {\sc{polpow*TBabs*zxipcf$_{1}$*zxipcf$_{1}$*diskbb}}. The right panels show the respective residuals of the best-fitting model to the data. The best-fitting parameters are given in Table \ref{spectra_table}.}
\label{spectra}
\end{figure*}
\section{Results}
\label{Sec_res}
We employed two different methods to determine the X-ray polarization of 4U 1630--47: (i) the model-independent {\sc{pcube}} method \citep{kis15}, and (ii) the spectro-polarimetric modeling of the \textit{I}, \textit{Q}, and \textit{U} spectra in {\sc{xspec}}. We describe the results with these two methods in \S~\ref{sec_pcube} and \S~\ref{spectro-pol}, respectively.

\subsection{Polarization measurements with \textit{IXPE}}
\label{sec_pcube}
Using the {\sc{pcube}} algorithm of {\sc{ixpeobssim}}, we measured a significant polarization for 4U 1630--47. We found a polarization degree PD $\sim 8$ \%
at a polarization angle PA $\sim 18^{\circ}$ in the $2-8$ keV band. In Figure \ref{2_8_stoke} we show the 4D contour plots of the polarization degree and polarization angle and the stokes parameters, $Q/I$ and $U/I$, in the $2-8$ keV band. Here, the contours are drawn at 68.27, 95.45, and 99.73 percent confidence levels for both Stokes parameters.
 
The values of all parameters for each detector unit are given in Table \ref{pol_table}.
Further, we extracted all parameters in four energy bands, $2.0-2.8$ keV, $2.8-4.0$ keV, $4.0-5.7$ keV, and $5.7-8$ keV and studied their variation with energy as shown in Figure \ref{ene_pol}. We observed that the PD increases by a factor $\sim 2$ from $2.0-2.8$ keV to $5.7-8$ keV (upper right panel of Figure \ref{ene_pol}) while the PA is consistent with being constant at 18$^{\circ}$ as a function of energy (upper left panel). The uncertainties in the measurement of the polarization angle in the 2--3 keV band is attributed to the low modulation factor ($\sim$ 13$\%$) of the three detector units (as shown in Table 4 and Figure 10 of \citealt{di22}). A similar increase in the PD as a function of energy is reported by \citet{kr22} for black hole binary source Cygnus X-1. As expected, the normalized stokes parameters, $Q/I$ and $U/I$, show a similar trend as the polarization degree.

\begin{table}
 \caption{The Polarization Degree (PD), polarization angle (PA) and normalized Stokes parameters of 4U 1630$-$47 extracted using the {\sc pcube} algorithm for the individual detector units, DU1, DU2 and DU3, and all DU's combined. }
 \begin{center}
\scalebox{0.95}{%
\hspace{-1.5cm}
\begin{tabular}{ccccc}
\hline
 & DU1 & DU2 & DU3 & All DUs\\ \hline
PD (\%) & $7.9\pm{0.3}$ & $8.2\pm{0.3}$ & $8.9\pm{0.3}$ & $8.3\pm{0.3}$\\
PA $(^\circ)$ & $16.0\pm{1.1}$ & $18.8\pm{1.1}$ & $18.5\pm{1.0}$ & $17.8\pm{1.1}$\\
$Q/I$ (\%) & $6.7\pm{0.3}$ & $6.5\pm{0.3}$ & $7.1\pm{0.3}$ & $6.8\pm{0.3}$\\
$U/I$ (\%) & $4.2\pm{0.3}$ & $5.0\pm{0.3}$ & $5.4\pm{0.3}$ & $4.9\pm{0.3}$\\
\hline
\end{tabular}}
\end{center}
\label{pol_table}
\end{table}
\subsection{Spectro-polarimetric analysis with \textit{NICER} and \textit{IXPE}}\label{spectro-pol}

To obtain a first estimate of the spectral properties of the source, we initially fitted only the {\it{NICER}} spectrum. Because the column density, $N_H$, towards the source is high \citep[$8-12\sim 10^{22}$ cm$^{-2}$,][]{mi15,wa16,ga19}, any emission below $\sim 1-2$ keV is likely due to calibration artefacts (C. Markwardt, \textit{NICER} calibration scientist, priv. comm.); therefore, we excluded the data below 2 keV, and fitted the \textit{NICER} spectrum in the $2-12$ keV band.
We fitted the spectrum with a single {\sc diskbb} component \citep{mi84} affected by interstellar absorption, {\sc{tbabs*diskbb}} in {\sc xspec}. For the {\sc tbabs} component we used the abundance tables of \citet{wi00} and the cross-section tables of \citet{ve96}. The fit gives $\chi^2=703.6$ for 151 degrees of freedom (dof), and the unabsorbed $2-12$ keV flux of the disk component is $\sim 1 \times 10^{-8}$ cm$^{-2}$ s$^{-1}$.
We added a power law to the model to account for the possible emission from the corona, but the fit does not improve and the power law is not significantly detected with a 95 \% upper limit of the flux in the $2-12$ keV band of $5.8\times10^{-11}$ erg cm$^{-2}$ s$^{-1}$ for an assumed power-law index $\Gamma=2.5$. Together with the absence of any significant variability in the PDS (see \S~\ref{sec_obs}), this confirms that the source was in the high-soft state during these observations.  \\

The high $\chi^2$ for the fit with a simple {\sc diskbb} model is due to the presence of multiple absorption lines in the $6-9$ kev energy range, with the most prominent ones at $\sim 6.70$ keV, 6.98 keV, 7.85 keV, 8.24 keV and 9.39 keV, possibly from iron and nickel, as reported previously by \citet{ku07}, \citet{ro14} and \citet{mi15}. We also observed residuals at $\sim 2.4$ keV that are likely instrumental from the gold M shell edges. We added the photo-ionized component {\sc{zxipcf}} to the model to account for the multiple iron absorption lines. Fitting the data with the model {\sc{TBabs*zxipcf*diskbb}} yields $\chi^2=237.6$ for 147 dof, but residuals at 6.98 keV remain. We added a second {\sc{zxipcf}} component to the model to account for this, possibly a blue-shifted iron line, and other residuals seen in the upper left panel of Figure \ref{spectra}. In the final fits we linked the ionization parameter and the column density of the two absorbers because, if we let them free, the two parameters are consistent with the respective parameters of the other ionized absorber within errors. The addition of this second {\sc zxipcf} component improves the fit significantly, with $\chi^2=173.3$ for 146 dof.

Even after using this model, there are still residuals in the $6-9$ keV energy range, as shown in the right panel of Figure \ref{spectra}. These residuals likely arise because the {\sc{zxipcf}} model does not allow to fit for variable abundances of the ionized material and assumes that the ionising source is a power law with index 2 whereas, in this case, the ionising source is a {\sc{diskbb}}. Since our purpose is to study the polarimetric properties of the source and not the physics of the wind, and because the reduced $\chi^2$ we obtain from the fit with this model is close to 1, we did not add other components to try and fit those residuals. 

Next, we carried out the spectro-polarimetric analysis fitting simultaneously the $Q$ and $U$ Stokes spectra of all detectors of \textit{IXPE} in the $2-8$ keV energy band and the {\it{NICER}} spectrum in the $2-12$ keV band in {\sc{xspec}}. We refrain from using the count spectrum ($I$) from \textit{IXPE} as the {\it{NICER}} spectrum can constrain the spectral model parameters better than the {\it IXPE} $I$ spectrum, thereby achieving finer information about the polarization parameters from the $Q$ and $U$ spectra. We first considered the model combination- {\sc{polconst*tbabs*zxipcf$_1$*zxipcf$_2$*diskbb}} where {\sc{polconst}} is a multiplicative model that assumes that the polarization of the source is energy-independent; the model has two free parameters, $A$, the polarization fraction and $\psi$, the polarization angle (in degrees). For a systematic error of 1.5 \%, the spectral fitting gives $\chi^{2} = 354.2$ for 224 dof. All the parameters were kept free during the fitting, with the common parameters of the model linked for the $Q$ and $U$ spectra of {\it IXPE} and the spectrum of {\it NICER}. 

Subsequently, we tried the polarization model {\sc{polpow}}, which considers a power-law dependence with energy of the polarization fraction and polarization angle. The model gives $\mathrm{PD}(E)=A_{\rm norm} \times E^{-A_{\rm index}}$ (in fractional units) and $\mathrm{PA}(E)=\psi_{\rm norm} \times E^{-\psi}_{\rm index}$ (in degrees). In the fits, we fixed the $\psi_{\rm index}$ to zero because, if we let that parameter free, the best-fitting value is 0.17$\pm$0.11, which is consistent with zero within errors; on the other hand, letting this parameter free does not improve the fits significantly. The fit with this model yields a $\chi^2= 269.1$ for 223 degrees of freedom. This is a significantly better fit than the one with {\sc polconst} with one less degree of freedom, with an F-test probability of $\sim 10^{-15}$. Figure \ref{spectra} shows the joint fits of the {\it{NICER}} count spectrum (top left panel), \textit{IXPE} $Q$ (middle left panel) and $U$ spectra (bottom left panel) for all three detector units, along with their residuals in the corresponding right panels. Table \ref{spectra_table} gives the best-fitting model parameters with $1\sigma$ error bars.  Integrating the model in the $2-8$ keV energy range we find PD $\sim 7.3$ \% and PA $\sim 18^\circ$. These values agree with those we obtained using the {\sc{pcube}} algorithm within error bars (see section \ref{sec_pcube}), which confirms the accuracy of the polarization measurements. Although the \textit{IXPE} energy band and energy resolution are limited, it should be noted that, because we fitted simultaneously the \textit{IXPE} and \textit{NICER} data with common parameters linked in {\sc xspec}, contrary to previous work we did not freeze any of the model parameters during the fits and could still accurately measure the polarization parameters from our fits.

\begin{table}
\centering
 \caption{{\it{NICER}}'s and {\it{IXPE's}} best-fitting spectral parameters for 4U 1630$-$47 with the model {\sc{polpow*TBabs*zxipcf$_1$*zxipcf$_2$*diskbb}}.}
 \begin{center}
\scalebox{0.95}{%
\hspace{-1.5cm}
\begin{tabular}{lll}
\hline

 Component    & Parameter &  Value\\ \hline
polpow  & $A_{\rm norm}$ ($10^{-2}$) &  $2.5 \pm{0.3}$\\
polpow  & $A_{\rm index}$ &  $-0.66 \pm {0.07}$\\
polpow  & $\psi_{\rm norm}$ (degree) & $17.9\pm{0.5}$\\
TBabs     & $N_H$ ($10^{22}$ cm$^{-2}$) &  $12.18\pm{0.04}$ \\
zxipcf$_{1}$  & $N_H$ ($10^{22}$ cm$^{-2}$) & $31^{+7}_{-4}$\\ 
zxipcf$_{1}$   & $\log{\xi}$ &  $4.73^{+0.05}_{-0.07}$\\
zxipcf$_{1}$   &Covering fraction &  $\le$0.93 \\
zxipcf$_{1}$   & Redshift ($10^{-3}$) &  $-6.3^{+0.8}_{-1.1}$\\
zxipcf$_{2}$   & Redshift ($10^{-3}$) &  $3.2^{+1.2}_{-0.9}$ \\
diskbb  & $T_{in}$ (keV)  & $1.472 \pm {0.003}$\\
diskbb   & norm &  $166\pm 2$ \\
diskbb & Flux$^{\dagger}$ ($10^{-8}$ erg cm$^{-2}$ s$^{-1}$) & $1.043\pm 0.003$ \\
\hline
$\chi^2$/dof & & 269.1 / 223 \\
\hline
\multicolumn{3}{l}{$^{\dagger}$ Unabsorbed flux in the $2-12$ keV range.}
\end{tabular}}
\end{center}
\label{spectra_table}
\end{table}
\section{Discussion and Conclusions}\label{sec_concl}
For the first time, we report a significant polarization degree of $\sim 8$ \% in the $2-8$ keV band with a polarization angle of $\sim 18^\circ$ for black hole binary source 4U 1630$-$47. The spectral study shows that the source is in the high soft state with negligible coronal emission, similar to what was observed in the 2006 and 2008 outburst of this source \citep{ca15}. The blue- and red-shifted absorption lines at $\sim 6.70$ keV, 6.98 keV, 7.85 keV, 8.24 keV and 9.39 keV in the NICER spectra provide a clear signature of the presence of disk winds \citep{ku07,ro14,mi15} during these observations.\\

On the basis of an outflowing corona model, \citet{po23} predicted that the PD could drop from 10 \% to 5 \% for high inclination ($i=60^\circ$) to low inclination  ($i=30^\circ$) systems. In the case of 4U 1630$-$47, however, the absence of any significant emission from a hard component (see \S\ref{spectro-pol}) excludes the possibility that the observed polarization is produced in an outflowing corona. We note also that there is no agreement about what the inclination angle of the source is. While \citet{ku98}, \citet{di13}, \citet{ki14}, and \citet{se14} reported that 4U 1630--47 is an edge-on source ($i \lessapprox 70{^\circ}$), \citet{ro14} proposed that it is a face-on system ($i \sim (11 \pm 5)^\circ$).\\~\\
Using {\it{PoGO+}} observations, \citet{ch18} reported that the polarization angle in Cyg X-1 is aligned with the radio jet. Later, \citet{kr22} reported similar results for Cyg X-1 using {\it IXPE} observations. Similarly, using {\it IXPE} data, \citet{lo22} and \citet{fa23} found the same for the neutron-star binary systems Sco X-1 and Cyg X-2. Since there are no measurements of the position angle of the radio jet for this source, we cannot compare our results with those other ones.
In the case of 4U 1630--47, however, it is very unlikely that the polarization is due to the corona \citep{lo22,fa23}, given that the corona is significantly absent in our observations. On the other hand, since in black-hole binary systems in the HSS the jets are quenched \citep[e.g.,][]{fe99}, the jet cannot be the source of polarization either in this case.\\

An interesting possibility is that the polarization in 4U 1630--47 is produced by the disk winds. We observed blue- ($z=-0.006$) and red-shifted ($z=0.003$) absorption lines in the {\it NICER} spectrum, corresponding to winds with velocities of $\sim 950-1900$ km s$^{-1}$. Using Monte Carlo ray-tracing techniques, \citet{sc09} showed that for black holes in the thermal-dominated state, if the polarization is a result of self-irradiation of the disk, the polarization angle should be perpendicular to the disk plane, so aligned with the jet axis (assuming that the jet is perpendicular to the disk plane). On the other hand, \citet{be98} solved the radiative transfer equations for a plane-parallel slab and found that, near Compton equilibrium with the radiation field, the wind produces polarization parallel to the disk plane, so perpendicular to the jet. Since the position angle of the jet in 4U 1630--47 is not known, we cannot test the predictions of \citet{sc09} and \cite{be98}. Using optical polarimetry, \citet{ko20} reported an increase in the PD with the appearance of winds in the black-hole binary MAXI J1820+070. \cite{ko20} further reported that, in that case, the polarization angle coincides with the angle of the jet. Based  on these results of the optical polarimetry for MAXI J1820+070 and the models of \cite{sc09} and \cite{be98}, we speculate that the position angle of the jet on the plane of the sky for 4U 1630$-$47 should be either $18^\circ$ or $108^\circ$.\\

As shown in Figure \ref{ene_pol}, the polarization fraction of 4U~1630$-$47 increases from $\sim$~6$\%$ at 2 keV to $\sim 11\%$ at 8 keV. Using numerical simulations and considering the direct and reflected emission component of the accretion disk, \citet{kr12} predicted a similar increase in the polarization fraction as a function of energy (in the $2-8$ keV band) for a black hole with a spin parameter $_{*}a \ge 0.9$ and an inclination angle to the line of sight of $75^\circ$. Later, considering the effect of absorption by the disk material, \citet{ta20} found  similar polarization spectra for the direct and reflected emission component for a highly spinning black hole ($a= 0.998$) with an inclination angle of the disk with respect to the line of sight in the range of 45 $^\circ-$ 75 $^\circ$ (see Figure 12 of \citealt{ta20}). Thus, assuming a highly spinning black hole in 4U 1630$-$47 \citep{se14}, and considering the disagreement of the inclination angle of the system \citep{ku98,di13,ki14,se14,ro14}, the most plausible scenario is that the observed polarization fraction of $\sim 8 \%$ in the 2--8 keV band is due to the thermal and/or reflected components.

\begin{acknowledgments}
We are grateful to an anonymous reviewer for their constructive comments, which helped us significantly improve the quality of the manuscript. This research has used data from the High Energy Astrophysics Science Archive Research Center Online Service, provided by the NASA/Goddard Space Flight Center. DR acknowledges Tomaso M. Belloni for providing the GHATS package used in this work for timing analysis. MM acknowledges support from the research program Athena with project number 184.034.002, which is (partly) financed by the Dutch Research Council (NWO). MM has benefited from discussions during Team Meetings of the International Space Science Institute (Bern), whose support he acknowledges.
\end{acknowledgments}
\vspace{5mm}
\appendix
\section{Table and Figures}
\begin{figure*}
\centering\includegraphics[scale=0.43,angle=0]{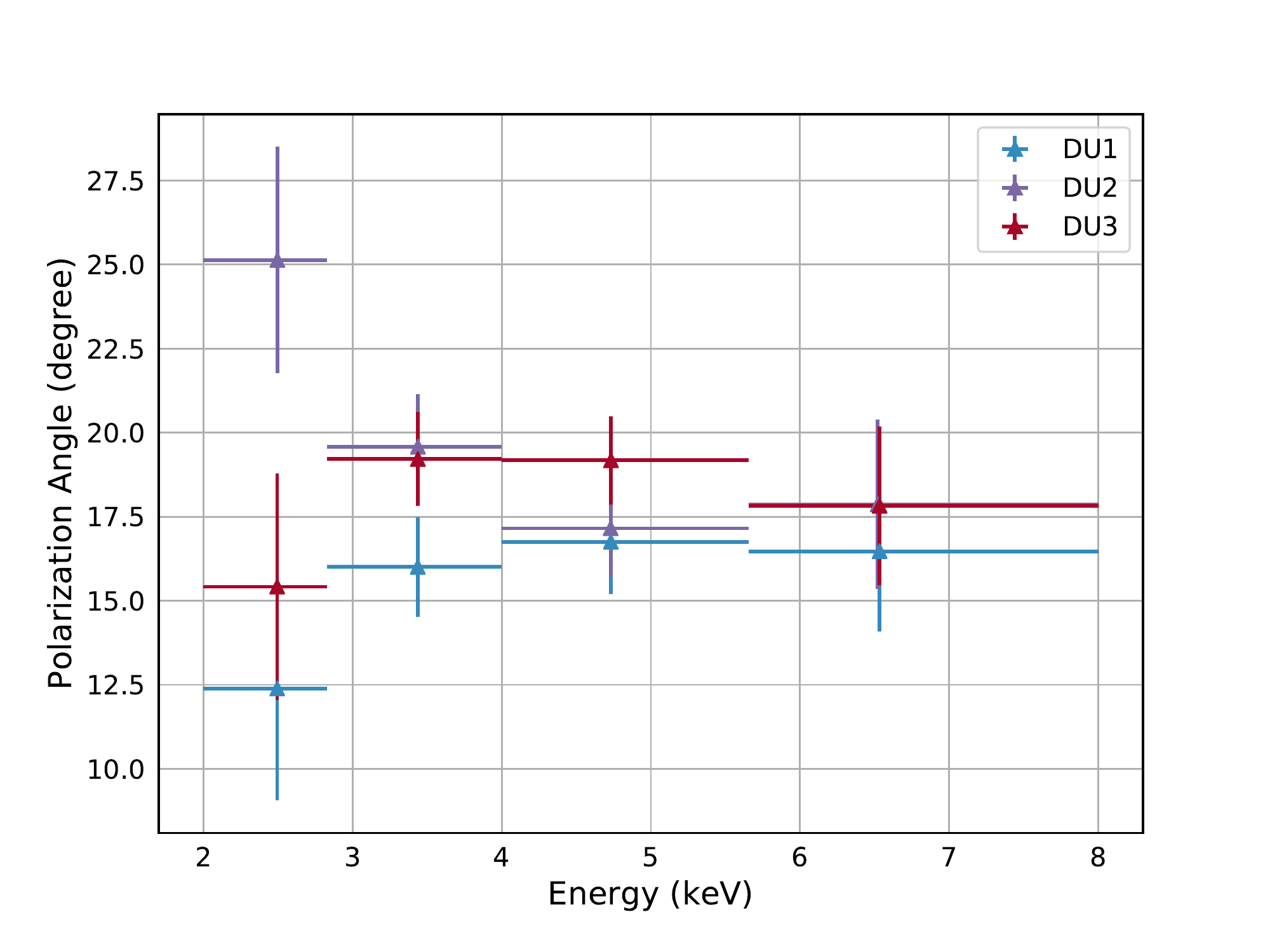}
\centering\includegraphics[scale=0.43,angle=0]{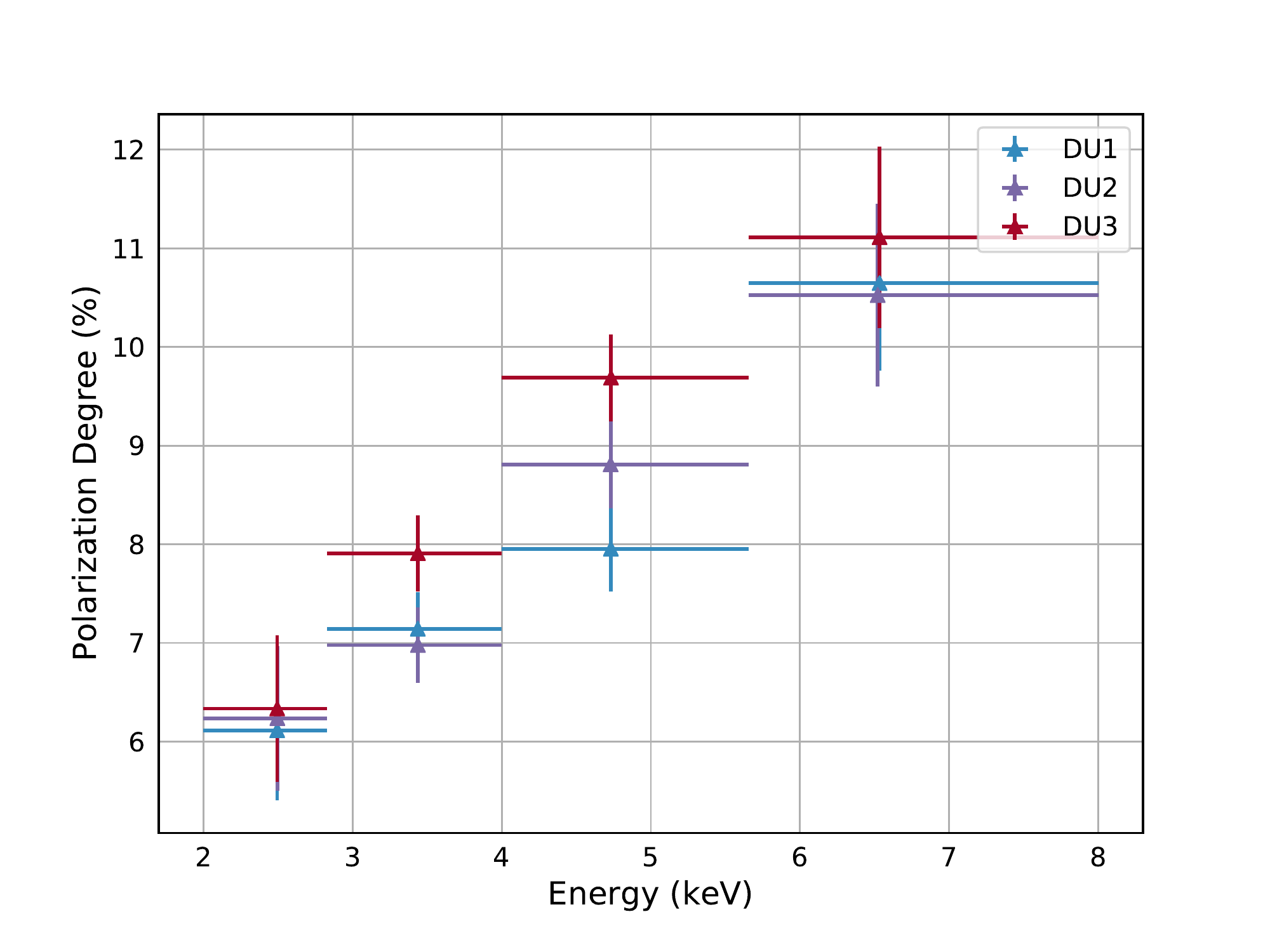}
\centering\includegraphics[scale=0.43,angle=0]{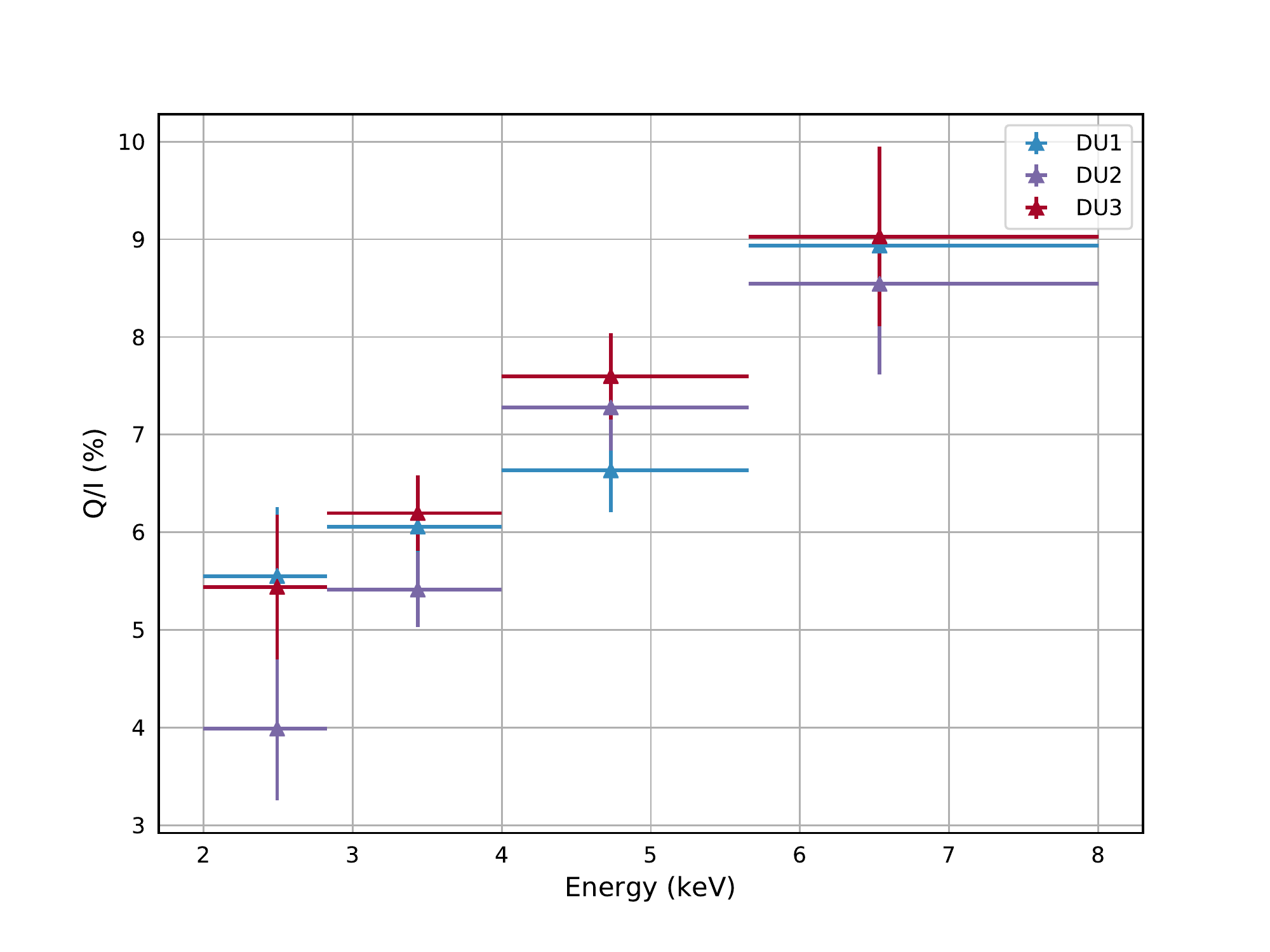}
\centering\includegraphics[scale=0.43,angle=0]{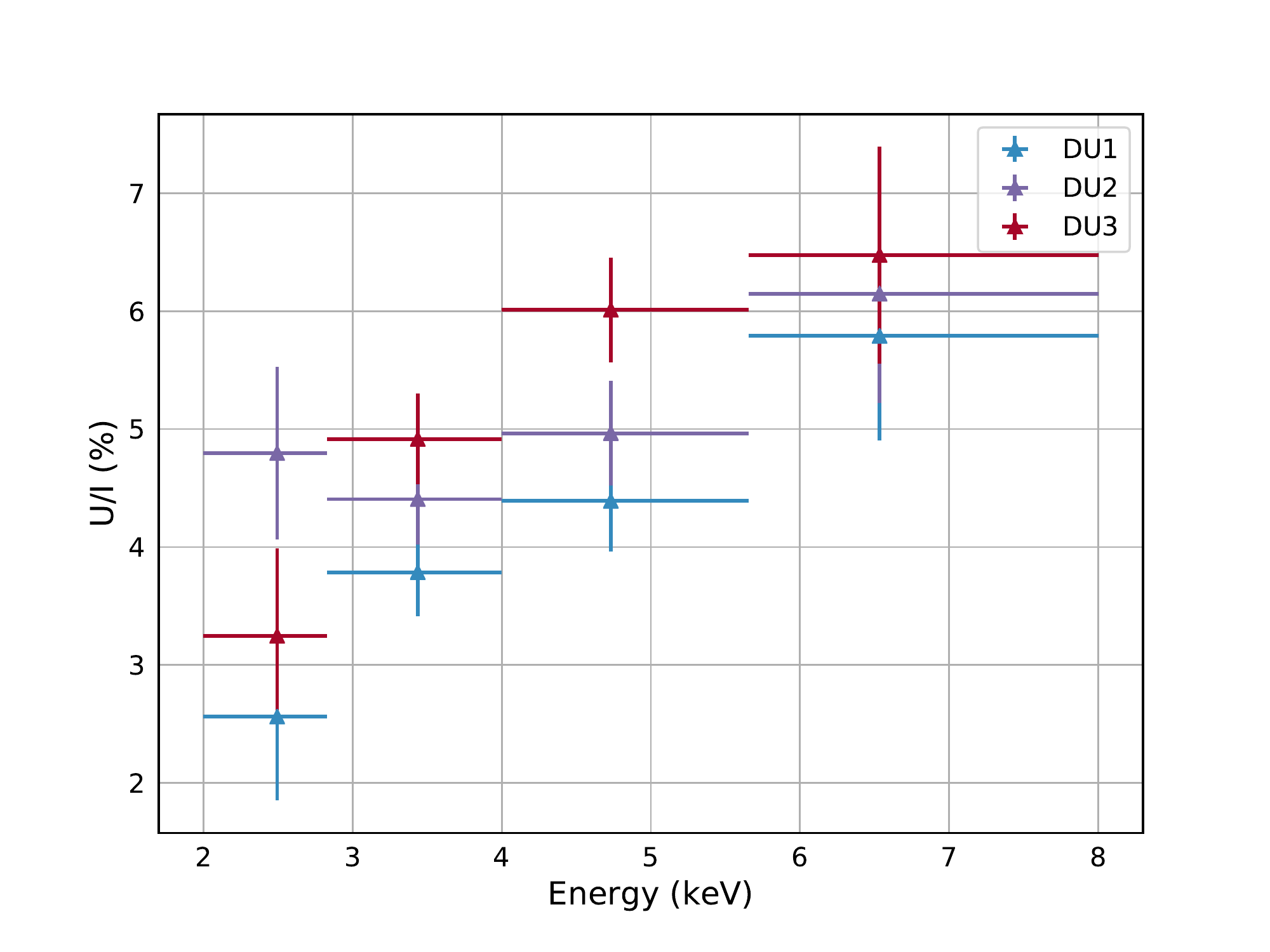}
\caption{The polarization angle, PA (upper left), polarization degree, PD (upper right), and normalized Stokes parameters, $Q/I$ (lower left) and $U/I$ (lower right), of 4U 1630$-$47 as a function of energy.}
\label{ene_pol}
\end{figure*}

In this Appendix we provide the observation log for 4U 1630$-$47 in Table \ref{obs_log}. We also show the energy dependence of the PA and PD in Figure \ref{ene_pol}.
\counterwithin{table}{section}

\begin{table*}
 \caption{Observation log for 4U 1630$-$47. The columns are the Instruments used, their ObsID, start and end date of the observation with exposure time.}
\scalebox{1.0}{%
\begin{tabular}{ccccc}
\hline
Instrument &ObsiD & Tstart & Tstop  & exposure \\
& & YYYY-MM-DD hh:mm:ss & YYYY-MM-DD hh:mm:ss & (seconds)\\ 
\hline
NICER  & 5501010102 &  2022-08-23 00:15:07 & 2022-08-23 23:39:40 & 1967\\
 & 5501010103 & 2022-08-24 01:01:43 & 2022-08-24 22:53:00 & 454\\
 & 5501010104 & 2022-08-25 01:42:36 & 2022-08-25 22:26:00 & 3920\\
 & 5501010105 & 2022-08-26 04:23:36 & 2022-08-26 21:40:20 & 2165\\
 & 5501010106 & 2022-08-27 03:36:17 & 2022-08-28 00:00:00 & 2152\\
 & 5501010107 & 2022-08-28 01:16:54 & 2022-08-28 15:29:00 & 2357\\
 & 5501010108 & 2022-08-29 06:43:00 & 2022-08-29 19:21:00 & 2973\\
 & 5501010109 & 2022-08-30 01:20:00 & 2022-08-30 23:12:20 & 904\\
 & 5501010110 & 2022-08-31 02:03:39 & 2022-08-31 22:24:40 & 3841\\
 & 5501010111 & 2022-09-01 01:27:00 & 2022-09-01 13:52:40 & 2556\\
\hline
IXPE & 01250401 & 2022-08-23 23:14:11 & 2022-09-02 18:54:11 & 458518 \\

\hline
\end{tabular}}
\label{obs_log}
\end{table*}
\counterwithin{figure}{section}

\bibliography{manuscript}{}
\bibliographystyle{aasjournal}

\end{document}